\documentclass[journal=acsnano,manuscript=article]{achemso}

\usepackage[version=3]{mhchem} 
\usepackage{multirow}
\usepackage{SIunits}

\author{Nathan Aubergier}
\affiliation{Laboratoire National des Champs Magn\'etiques Intenses, CNRS, LNCMI, Universit\'e Grenoble Alpes, Univ Toulouse 3, INSA Toulouse, EMFL, F-38042 Grenoble, France}
\alsoaffiliation{Universit\'e Grenoble Alpes, CEA, Grenoble INP, IRIG, Pheliqs, 38000 Grenoble, France}
\author{Vincent T. Renard}
\affiliation{Universit\'e Grenoble Alpes, CEA, Grenoble INP, IRIG, Pheliqs, 38000 Grenoble, France}
\author{Sylvain Barraud}
\affiliation{Universit\'e Grenoble Alpes, CEA-LETI, 38000 Grenoble, France}
\author{Kei Takashina}
\affiliation{Department of Physics, University of Bath, Bath, BA2 7AY, UK}
\author{Benjamin A. Piot}
\email{benjamin.piot@lncmi.cnrs.fr} \affiliation{Laboratoire National des Champs Magn\'etiques Intenses, CNRS, LNCMI, Universit\'e Grenoble Alpes, Univ Toulouse 3, INSA Toulouse, EMFL, F-38042 Grenoble, France}

\title[Wide Electrical Tunability of the Valley Splitting in a Doubly gated Silicon-on-Insulator Quantum Well]
  {Wide Electrical Tunability of the Valley Splitting in a Doubly gated Silicon-on-Insulator Quantum Well}

\begin{document}

%
%
%
%

\begin{abstract}

 The valley splitting of 2D electrons in doubly-gated silicon-on-insulator quantum wells is studied by low temperature transport measurements under magnetic fields. At the buried thermal-oxide SiO$_{2}$ interface, the valley splitting increases as a function of the electrostatic bias $\delta n = n_{B}-n_{F}$ (where $n_{B}$ and $n_{F}$ are electron densities contributed by back and front gates, respectively) and reaches values as high as $6.3$~meV, independent of the total carrier concentration of the channel. We show that $\delta n$ tunes the square of the wave function modulus at the interface and its penetration into the barrier, both of which are key quantities in a theory describing interface-induced valley splitting, and is therefore the natural experimental parameter to manipulate valleys in 2D silicon systems. At the front interface, made of a thin ``high-k'' dielectric, a smaller valley splitting is observed, adding further options to tune the valley splitting within a single device.
\end{abstract}



The valley in which electrons reside in the momentum space of the Brillouin zone (BZ) is a pivotal quantum degree of freedom in solid-state physics. While it can potentially be used in a similar fashion as spins are used to convey quantum information, within the emerging field of ``valleytronics'', a valley degeneracy can also be an undesirable source of quantum decoherence in low dimensional spin-based quantum-bits (qubits) systems.
Silicon, which is both the pillar of modern electronics and a suitable material for the realization of spin qubits\cite{Stano2022,Burkard2023}, has a six-fold bulk valley degeneracy in the conduction band with degenerate energy minima lying close to the six X points of the BZ. In electronic systems confined into a plane perpendicular to the [100] direction, the four in-plane valleys are pushed to energies higher than the two out-of-plane valleys because of anisotropy in the effective mass. The six-fold bulk (3D) degeneracy is thus reduced to two in [100] silicon quantum wells.
 In spite of the ubiquitous use of 2D silicon, our understanding of how the degeneracy of these two valley energy levels can be lifted in actual devices is still uncomplete. Early theories\cite{Ohkawa1977-III} have pointed out that intervalley scattering (leading to ``valley splitting'') can occur in the presence of an abrupt interface. The recent progress in silicon-based qubits have motivated new experimental\cite{PaqueletWuetz2022,Losert2023} and theoretical\cite{Feng2022,Losert2023,Adelsberger2024} studies to better characterize the valley splitting in silicon and propose routes for its optimization, meaning, in the context of spin-based qubits, its maximization. The highest mobility silicon systems generally realized at the Si/SiGe interface unfortunately exhibit rather small valley splittings (the order of a few to hundreds of $\mu$eV) with strong variations between structures, while conventional metal-oxide-semiconductor-field-effect-transistors (MOSFETs)or silicon-on-insulator systems, of lower mobilities, generally display larger valley splittings (up to tens of meV). Experimental observations, together with more recent theoretical works \cite{Friesen2007,Saraiva2009} highlight the crucial role of the nature of the confining interface. In addition, measurements in singly-gated MOSFETs systems have demonstrated that the valley splitting is increasing with increasing densities of carriers \cite{Nicholas1980ssc,Lodari2022}, consistent with a role played by the (density-induced) electric field at the interface \cite{Ohkawa1977-III}. In doubly-gated systems with the ``SIMOX'' type of buried SiO$_{2}$ oxides\cite{Takashina2004,Takashina2006}, the manipulation of the electric field along the confinement direction can generate valley splittings as high as tens of meV. While such observations can be qualitatively foreseen with the theory of the electric breakthrough (where the electric field tends to push the electronic wave function into the barrier and enhance the role of the interface), the absence of a systematic disentanglement between vertical electric fields and carrier density and the remaining quantitative variations from one system to another call for further characterization of silicon devices displaying different interfaces in the presence of variable (and controlled) electric fields.

In this Letter, we present a study of the valley splitting in doubly-gated silicon-on insulator quantum wells using low temperature transport measurements under magnetic and electric fields. The double gate configuration allows for a disentanglement of effects related to the electron density, on the one hand, and to electrostatic bias $\delta n = n_{B}-n_{F}$ where $n_{B}$ ($n_{F}$) is the contribution from the back (front) gate to the total density, on the other hand. This latter is found to be the key parameter driving the strength of the valley splitting, which can reach value as high as $6.3$~meV as electrons are pushed toward the bottom SiO$_{2}$ thermal oxide of our device, independent from the channel total carrier concentration $n = n_{B}+n_{F}$. More precisely, we show that the wave function modulus at the interface and its penetration within the barrier, which are the relevant quantities in the theoretical description of interface-induced valley splitting made by Saraiva \textit{et al}\cite{Saraiva2009}, are both proportional to the electrostatic bias $\delta n$ and independent of $n$. The role of the nature of the interface in determining the valley splitting is confirmed by the lower splitting observed at the front interface of our device made out of a thin ``high-k'' dielectric. The interface and electric field-dependent valley splitting therefore result in a wide tunability within a single device, opening new perspectives to engineer valley-tunable devices.

\begin{figure}
\includegraphics[width=1\linewidth,angle=0,clip]{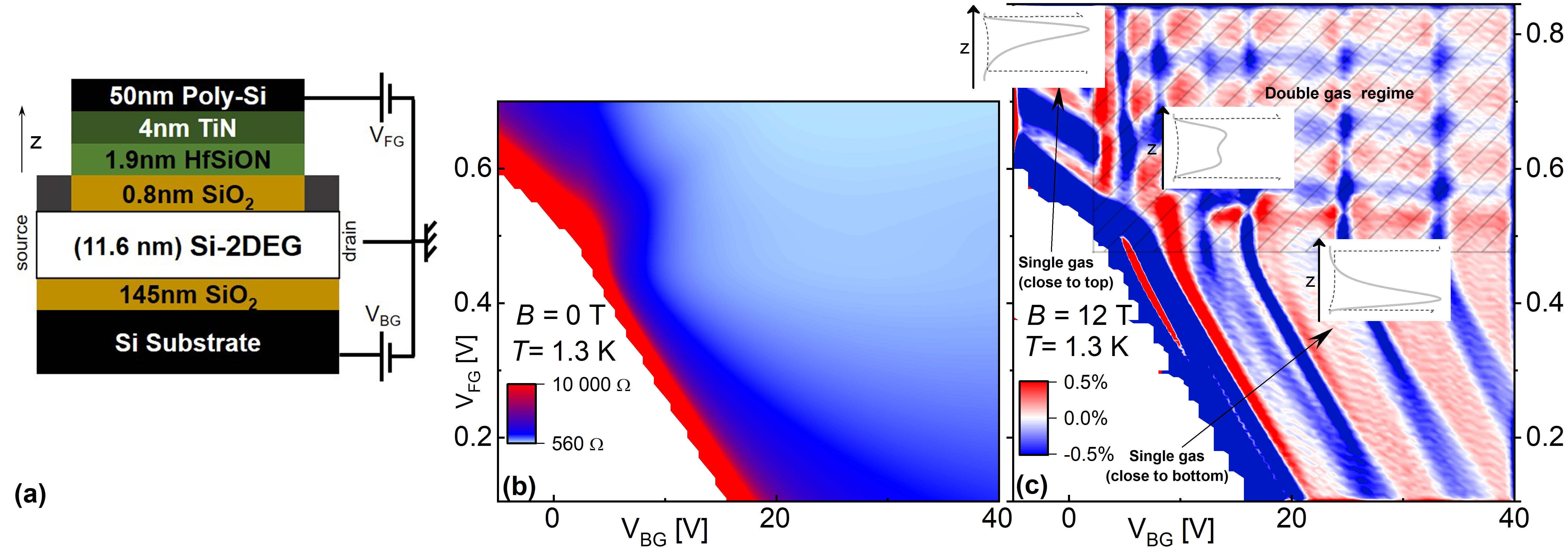}
\caption{a) Schematical structure of the doubly-gated silicon transistor studied in this work. The typical thickness of each layer is specified. b) Typical $B=0$~T resistance color map as a function of the front and back gate voltages, respectively $V_{FG}$ and $V_{BG}$. Light blue: +560$\Omega$, Red: +10k$\Omega$. c) Typical $B_{\perp}=$12~T resistance modulations due to quantum oscillations, revealing different conduction regimes (single gas close to the top interface, single gas close to the bottom interface, and double gas). Red: +0.5 \%, white: 0\% and blue: -0.5\%.
Insets: simulations of the typical confinement potential $V_{QW}$ (short-dashed black line, top axis) and electronic density distribution (light grey line, bottom axis) along the confinement axis (z) for each conduction regime, for a total electron density $n=4\times 10^{12}$~cm$^{-2}$. Insets are positioned approximatively at the ($V_{FG}$,$V_{BG}$) coordinates where calculations are performed.}\label{sample structure-fig}
\end{figure}

The sample studied here are doubly gated silicon MOSFETs (Metal-Oxide-Field-Effect-Transistor) schematically presented in figure~\ref{sample structure-fig}(a).
Both the front and back gate voltages ($V_{FG}$ and $V_{BG}$) can contribute to the total electron density $n$, which can be tuned between $4 \times 10^{11}$~cm$^{-2}$ and $ \sim 8\times 10^{12}$~cm$^{-2}$, the peak electron mobility being about $\sim3500$~cm$^{2}$V$^{-1}$s$^{-1}$. These are rather low values compared to state-of the art mobilities Si MOSFET\cite{Camenzind2021,Lodari2022}, due to the use of thin HfSiON high-k dielectric at the top of the structure granting the compactness/better operability of each transistor incorporated in electronic devices. In spite of this modest mobility, these devices have the advantage of enabling us to precisely control the vertical electric field independently from the electron density, at variance with the most of widely studied singly-gated systems.

A typical resistivity map as a function of the front ($V_{FG}$) and back ($V_{BG}$) gate voltage is shown in figure~\ref{sample structure-fig}(b). When both gate voltages are large, a so-called ``double gas'' regime with two maxima in the electron density vertical profile (see insets in figure~\ref{sample structure-fig}(c)) can be reached. In this case, the gas close to the top (bottom) interface is only controlled by the front (back) gate, which solely drives Shubnikov-de Haas oscillations in the dashed region of figure~\ref{sample structure-fig}(c) (see also Ref.\cite{takashina2006a}). When gated asymmetrically (i.e. with $n_{F}$ and $n_{B}$ being significantly different), the electronic wave function exhibits a single maximum located either near the front interface with the high-k dielectric, or the back Si/SiO$_{2}$ interface (a more conventional thermal oxide). Increasing asymmetry results in electrons being pressed further against the corresponding interface, providing us with the opportunity to study the valley splitting not only in the presence of electric fields of various magnitudes, but also at two very different interfaces.

\begin{figure}
\includegraphics[width=1\linewidth,angle=0,clip]{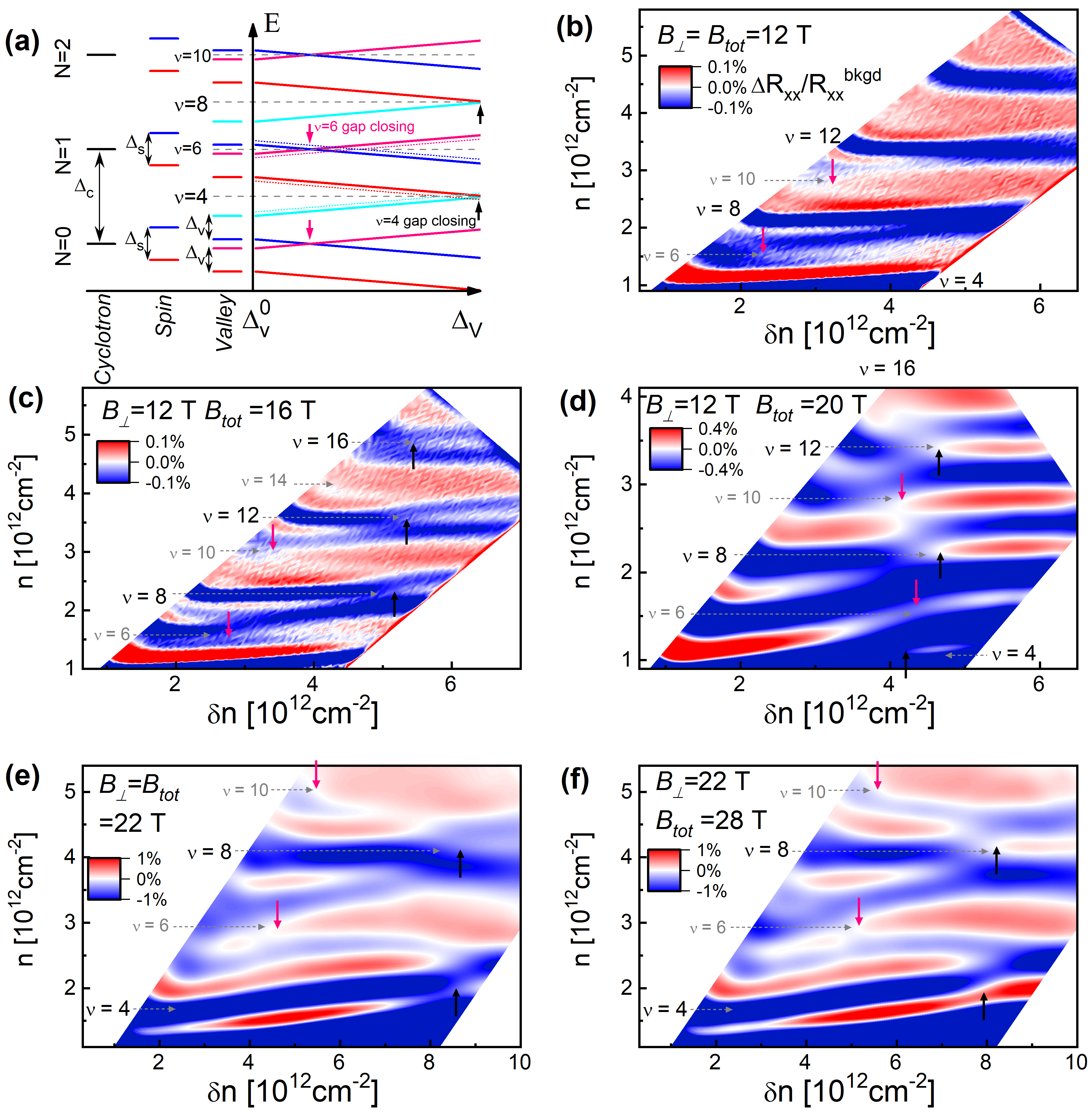}
\caption{a) Schematics of energy levels in the quantum Hall regime. $\Delta_{c}$ is the cyclotron gap, $\Delta_{s}$ is the spin gap and $\Delta_{v}$ is the valley gap. In our samples for, high enough normal magnetic fields $B_{\perp}$, $\Delta_{c} > \Delta_{s} > \Delta_{v}^{0}$, where $\Delta_{v}^{0}$ is the value of $\Delta_{v}$ at zero electrostatic bias ($\delta n = n_{B}-n_{F}$=0). Evolution for increasing valley splitting $\Delta_{v} > \Delta_{v}^{0}$. Short-dotted lines correspond to a higher value of the Zeeman energy than for solid lines. Down (up) arrows signal coincidence occurring at $\nu=4N+2$ ($\nu=4(N+1)$). (b-f) Longitudinal resistance modulations $\Delta R_{xx}/R_{xx}^{bkgd}$ (where $R_{xx}^{bkgd}$ is the background value) due to quantum oscillations in the $(n,\delta n)$ phase space, in the single gas regime, for different values of the perpendicular ($B_{\perp}$) and total ($B_{tot}$) magnetic fields. (b) are the same data as the one presented in figure~\ref{sample structure-fig}(c). $T=1.3$~K. Filling factors are indicated (sometimes with additional horizontal dashed arrows), and vertical down (up) arrows highlight precoincidence occurring at $\nu=4N+2$ ($\nu=4(N+1)$). For those
fixed integer filling factors, precoincidence are identified when $\Delta R_{xx}/R_{xx}^{bkgd}=0\%$, coded in white color.}\label{coincidences-fig}
\end{figure}

Pioneering measurements of the valley splitting in silicon 2D electron gas (2DEG) were performed using transport techniques in the quantum Hall regime, i.e. in the presence of a magnetic field perpendicular to the inversion layer \cite{Nicholas1980ssc}. In this regime, the electron orbital motion is quantized and the density of states is discretized into a set of Landau levels (LL) separated by the cyclotron gap $\Delta_{c}$, and the four-fold (spin and valley) total degeneracy leads to internal splitting of each LLs, as depicted in figure~\ref{coincidences-fig}(a).
A well-established approach to determine energy gaps in this regime, which we have focused on in this work, is the so-called ``coincidence'' method\cite{Fang1968,Nicholas1980ssc,Takashina2004} where the principle is to bring different LLs into coincidence in the energy spectrum by acting on external parameters. Coinciding levels generate local density of states maxima which can be probed through the longitudinal resistivity, and the coincidence condition establishes a relation between the different energy gaps involved in the LL spectrum. The gap of interest (here, the valley gap $\Delta_{v}$) can then be estimated from the other energy gaps (here, $\Delta_{c}$ and/or the spin gap $\Delta_{s}$) which are independently characterized. In our experiments, $\Delta_{c}$ is fixed by the component of the magnetic field perpendicular to the sample $B_{\perp}$, while the Zeeman energy (contributing to the spin gap) is determined by the total magnetic field $B_{tot}$, and the ratio between them can be modified by using a tilted magnetic field configuration. As we will show, the valley splitting can be tuned in our device by the sign and magnitude of the electrostatic bias $\delta n = n_{B}-n_{F}$, similarly to what was done previously in other doubly-gated systems with ``SIMOX'' type of buried SiO$_{2}$ oxides
\cite{Takashina2006,Renard2015}. In this context, the ability to tune $B_{\perp}$,  $B_{tot}$ and $\delta n$ over a wide range of values makes it possible to measure large variations of the valley gap via the coincidence method.

In figure~\ref{coincidences-fig}(b-f), we report longitudinal resistance variations with respect to its background value as a function of the electron density and the electrostatic bias $\delta n$, in the presence of magnetic fields. The experimental $(V_{FG},V_{BG})$ phase space here has been converted into a $(n,\delta n)$ phase space (see section II.C. of the supporting information), where one can disentangle both parameters to independently track the effect of each one. Filling factors $\nu=n/(eB_{\perp}/h)$, corresponding to the number of occupied LL (including spin and valley degeneracy), are indicated. Resistance minima associated with the usual cyclotron gaps at filling factors $\nu=4(N+1)$ (where $N\geq0$ is the Landau level index) can be observed, together with additional minima at filling factors $\nu=4N+2$ or $\nu=4N+2 \pm1$, which correspond for small values of $\Delta_{v}$ to the lifting of the spin and valley degeneracies, respectively.
The resistance minima observed at $\nu=4N+2$ ($N=1,2,3$) disappear above a critical value of $\delta n$ (highlighted by down arrows), signaling the closing of the spin gap expected when $\Delta_{v}$ reaches a sufficient value (see figure~\ref{coincidences-fig}(a)). In a given LL of index $N$, sublevels with different spin and valley coincides when $\Delta_{v}=\Delta_{s}$, as previously observed e.g. at $\nu=6$ \cite{Takashina2004}. When increasing $\delta n$ further, gap closing between LL of \textit{different} indexes can even be observed at $\nu=4(N+1)$ (with $N=1,2,3$) in figure~\ref{coincidences-fig}(c-f).
This now corresponds to the closing of the corresponding energy gap $\Delta^{\nu=4(N+1)}=\Delta_{c}-\Delta_{s}-\Delta_{v}=0$, which can only occur for much larger values of $\Delta_{v}$.
For a fixed and purely perpendicular magnetic field $B_{\perp}$, both $\Delta_{c}$ and $\Delta_{s}$ are set, and there is a single value of $\Delta_{v}$ which is reached at high enough $\delta n$ to  satisfy this latter equation. By tiling the sample in the magnetic field, for the same electron density $n$ and perpendicular field $B_{\perp}$,  a different value of the total field and thus of the Zeeman energy can be obtained, shifting the coincidence to a different value of $\Delta_{v}$ (see the energy diagram in figure~\ref{coincidences-fig}(a)), and thus a different $\delta n$. This is clearly observed as in figure~\ref{coincidences-fig}(b-d), where higher Zeeman energies precipitate gap closing to lower $\delta n$ for $\nu=4,8,12$ (see the vertical up arrows shifting toward lower $\delta n$ values between (c) and (d)). For the same reasons, gap closings at $\nu=6$ are delayed, as experimentally observed with vertical down arrows shifting toward higher $\delta n$ values between (b), (c) and (d). Ultimately in figure~\ref{coincidences-fig}(d), both types of coincidences (intra and inter LL) occur in a close $\delta n$ range. Experiments in higher magnetic fields (see e.g. figure~\ref{coincidences-fig}(e-f)) boost the cyclotron and spin gaps which respectively push the inter and intra LL coincidences to higher $\Delta_{v}$ ($\delta n$) values.

We now turn to quantitative extraction of $\Delta_{v}$. While the coincidence equations presented above describe a  ``spectral'' coincidence of each energy level, the presence of disorder, which can be significant in some region of our ($n,\delta n$) space, leads to an enhanced LL broadening so that LL overlap (and the resistivity value at $\nu=4(N+1)$ changes) before they actually reach the same energy. Additionally, the increasing disorder as the wave function is pushed toward the interface (with increasing $\delta n$), makes it nontrivial to locate the exact ``spectral'' coincidence which should be signaled by a local density of states maximum, followed by a decrease of the resistance after the LL crossing (see for example \cite{Takashina2006}). For this reason, we decided to focus on the situation when LL just starts to overlap (referred to as ``precoincidence'') which with a precise characterization of disorder enables us to extract $\Delta_{v}$ (the full procedure is described in section III.B. of the supporting information). This approach has the advantage to avoid a significant LL overlap which could result in modifying the spin and/or valley polarizations, and therefore potentially affect the value of the spin and valley splitting via many-body effects \cite{Piot2005,Khrapai2003} (see also section III.A. of the supporting information).
\begin{figure}
\includegraphics[width=1\linewidth,angle=0,clip]{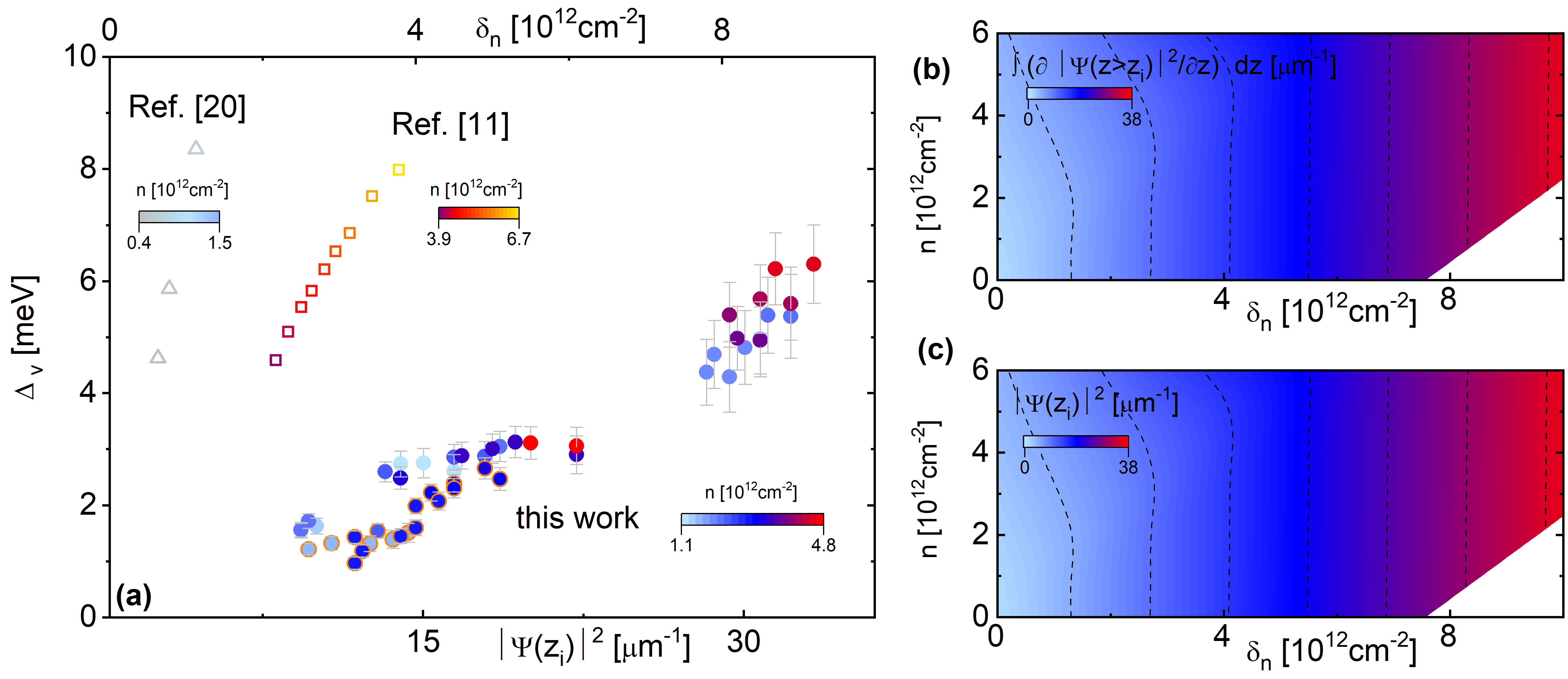}
\caption{a) Valley splitting $\Delta_{V}$  as a function of the electrostatic bias $\delta n$ (top axis) or the electronic wave function modulus at the interface $\mid\Psi(z=z_{i})\mid^{2} $ (bottom axis, calculated from simulations). Our samples are displayed as closed circles with color-coded electron density $n$. Data obtained from intra LL coincidences are marked with an additional orange edge. Data obtained for other silicon 2DEGs in the literature are shown with open symbols. For the latter, only the bottom x-axis scale ($\mid\Psi(z=z_{i})\mid^{2}$) is relevant in this plot. Data of Ref.\cite{Takashina2011} are obtained from Shubnikov-de Haas oscillations analysis with a procedure given in Ref.\cite{Takashina2006} and extend to higher $\Delta_{V}$ values (not shown). b) Calculations of $\mid\Psi(z=z_{i})\mid^{2}$  (b) and  $\int \Psi(z>z_{i})^{2}dz$ (c) vs $\delta n$  in our samples. Contour lines at equidistant values are plotted for each parameter as black dashed-lines. The region where those lines are no longer vertical corresponds to the double-gas regime.} \label{Fig3}
\end{figure}
In figure~\ref{Fig3}(a), we report the values of $\Delta_{V}$ obtained from 50 analysed precoincidences as a function of $\delta n$ (top axis). A general increasing trend of $\Delta_{V}$ vs $\delta n$ is observed, with an approximatively linear behaviour at low $\delta n$ and values as high as $6.3$~meV reached for the highest $\delta n \simeq 9.2$ studied. Thanks to the ability to tune the Zeeman/cyclotron ratio with tilted-field experiments, it is possible to extract different values of $\Delta_{V}$ for the \textit{same} electron density but different $\delta n$ values, which demonstrates that the parameter controlling valley splitting is $\delta n$, independent of the value of the carrier concentration. This is confirmed by the weak variation of $\Delta_{V}$ at fixed $\delta n$ as the carrier concentration is varied, which can be seen via the color-coding of each symbol. Points of different color (density) yield very similar $\Delta_{V}$ for example at $\delta n \simeq 3.8\times 10^{12}~cm^{-2}$, $\delta n \simeq 6\times 10^{12}~cm^{-2}$ or $\delta n \simeq 8.2\times 10^{12}~cm^{-2}$.

While $\delta n$ is related to the presence of a nonzero vertical electric field in the structure, it is informative to study how this experimental parameter is connected to the local properties of the electronic wave function close to the SiO$_{2}$ barrier, which are central in the theoretical descriptions of interface induced valley splitting \cite{Friesen2007,Saraiva2009}. In Ref.\cite{Saraiva2009}, the main effect of the vertical electric field is to modify the electronic probability at the Si/SiO$_{2}$ interface (quantified by the wave function modulus $\mid\Psi(z_{i})\mid^{2}$), where $z_{i}$ is the coordinate of the interface position,  and the spreading of the wave function within the SiO$_{2}$ barrier $\int \mid\Psi(z>z_{i})^{2}\mid dz$. Both quantities can be calculated for our sample structure by using Poisson-Schrodinger simulations (see section IV of the supporting information) and are shown in figure~\ref{Fig3}(b) and (c). The main observations are that, in the single gas regime, they both increase linearly as a function of $\delta n$ and are independent of the electron density $n$ (unlike the vertical electric field). This is fully consistent with the fact our experimental knob $\delta n$ pushes the electronic wave function into the SiO$_{2}$ barrier, which enhances these quantities and boosts the valley splitting, since a larger proportion of electrons can experience intervalley scattering \cite{Saraiva2009}. This confirms that $\delta n$ is a proper experimental parameter to tune the valley splitting, because of its connection to $\mid\Psi(z_{i})\mid^{2}$ and $\int \mid\Psi(z>z_{i})^{2}\mid dz$, which should be used as the relevant quantities to compare valley splitting in different systems. To do so, we have therefore evaluated these quantities for different silicon 2DEGs for which valley splitting measurements are available \cite{Takashina2011,Lodari2022}, and report in figure~\ref{Fig3}(a) the measured valley splittings as a function of $\mid\Psi(z_{i})\mid^{2}$.

We observe that all measured splittings increase approximately linearly with $\mid\Psi(z_{i})\mid^{2}$, confirming the universal connection between the magnitude of the valley splitting and the electronic wave function at the barrier interface. In different structures, different experimental parameters can modify this latter: in first approximation, the electron density for singly-gated MOSFETS (see the clear correlation between $\Delta_{V}$ and the symbol color for the data of  Ref.\cite{Lodari2022}), and $\delta n$ for doubly-gated systems such as ours or the ones in Ref.\cite{Takashina2011}. We note that similar values $\Delta_{V}$ can be observed in samples with mobility differing by more than one order of magnitude, showing that $\Delta_{V}$ is not sensitive to the raw amount of disorder (in our case, valley splittings of few meVs can still be observed for mobilities as low as $\sim400$~cm$^{2}$V$^{-1}$s$^{-1}$).
Different ``slopes'' for $\Delta_{V}$($\mid\Psi(z_{i})\mid^{2}$) are nevertheless observed for different samples, which we attribute to the different microscopic nature of interfaces in each cases. Indeed, the expected theoretical valley splitting has been shown to depend on the nature of the interface (material, barrier height) and to be very sensitive to  its microscopic profile, which can vary upon interface roughness or the type of scatterers present in its vicinity\cite{Saraiva2009,PaqueletWuetz2022}. In the ``SIMOX'' structures\cite{Takashina2004}, the effect of shear strain \cite{Adelsberger2024,Noborisaka2024} has also been recently put forward to account for the observed giant valley splitting. These effects will modify not only the bare valley splitting (i.e. in absence of any applied electric field), but also its dependence on $\mid\Psi(z_{i})\mid^{2}$, with different ``slopes'' expected for $\Delta_{V}$($\mid\Psi(z_{i})\mid^{2}$).

The influence of the interface nature on the valley splitting can also be observed by pushing electrons toward different interfaces in a single device\cite{Takashina2004}. In our device, one can explore the interface-induced valley splitting at the top interface, made of a high-k dielectric, by gating the 2DEG close to this latter (see figure~\ref{sample structure-fig}(c)). The valley splitting was there found to lie in the range $\sim1-2$~meV (see section V of the supporting information) for $\mid\Psi(z_{i})\mid^{2}\sim 17-22~\mu m^{-1}$, smaller than the one observed near the buried thermal SiO$_{2}$ oxide at similar $\mid\Psi(z_{i})\mid^{2}$. This could be related to the microscopic difference between the interfaces, the former involving a thinner barrier exhibiting higher electric disorder which affects the sharpness of the potential barrier at the interface, leading to smaller valley splittings \cite{Friesen2007,Saraiva2009}.

\begin{figure}
\includegraphics[width=1\linewidth,angle=0,clip]{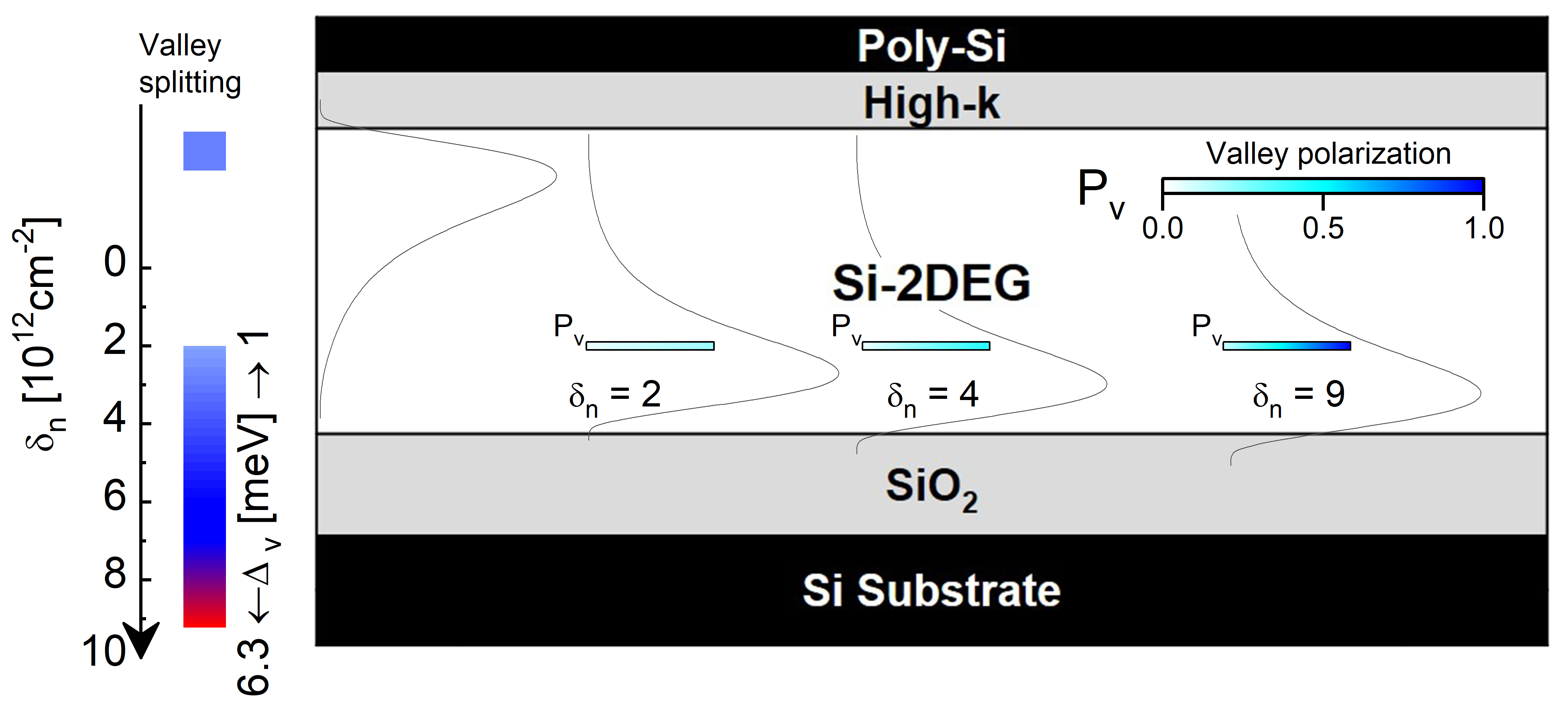}
\caption{a) Valley splitting tunability of our doubly-gated devices. The vertical colored bar on the left summarizes the reachable amplitudes of the valley splitting $\Delta_{V}$ in our devices.
At the bottom SiO$_{2}$ oxide interface, $\Delta_{V}$ can be tuned depending on the value of $\delta n$ (left vertical axis). At the top high-k interface, smaller $\Delta_{V}$ values are observed for similar conditions of wave function penetration into the barrier. Schematical electronic wave function vertical profiles (grey lines) at the interfaces for different representative values of $\delta n$ (indicated in units of $10^{12}~cm^{-2}$).
 The wave function at the top interface is schematically placed to correspond to the investigated experimental range $\mid\Psi(z_{i})\mid^{2}\sim 17-22~\mu m^{-1}$ at this interface. The 2DEG valley polarization $P_{v}=\Delta_{V}/(n \pi \hbar^{2}/m^{*})$ ranges accessible for a given $\delta n$, determined by the lower (conduction threshold) and upper (double gas regime formation) limits of $n$, are indicated with horizontal color bars within each wave function profile.}\label{Fig4}
\end{figure}

Figure~\ref{Fig4} summarizes the possibilities of tuning the valley splitting in our doubly-gated silicon transistors. As one can see, pushing the electron gas toward the bottom interface offers a wide tunability as a function of the electrostatic bias $\delta n$, while the valley splitting is smaller close to the top interface for a similar penetration into the barrier. One can approximate the degree of valley polarization $P_{v}$ of our 2D electron gas by considering the ratio $\Delta_{V}/2E_{F}$, where $E_{F}=(n \pi \hbar^{2})/(2 m^{*})$ is the density-dependent Fermi energy of the spin/valley degenerate 2DEG.
Thanks to the combination of a large sweepable ($n,\delta n$) phase space and interfaces of different nature, the studied devices offer a large control on the valley polarization of charge carriers at low temperatures. The existence of a double gas conduction regime also offers the interesting perspective to have two spatially distinct 2DEGs of various valley polarizations, which could potentially be electrically isolated to design devices such as valley polarization splitters.

\begin{acknowledgement}

We thank Y.M. Niquet, M Cass\'e, and M.O. Georbig for stimulating discussions. This work was supported by the Laboratoire d`Excellence LANEF (Grant No. ANR-10-LABX-51-01). We acknowledge the support of the LNCMI-EMFL, CNRS, Univ. Grenoble Alpes, INSA-T, UPS, Grenoble, France.

\end{acknowledgement}

\begin{suppinfo}

Additional data and analysis are available in the supporting information available online (https://pubs.acs.org/doi/10.1021/acs.nanolett.5c03049).
They include details on experimental setup, sample characterization, analysis procedures as well as additional experimental data.
\end{suppinfo}


\providecommand{\latin}[1]{#1}
\makeatletter
\providecommand{\doi}
  {\begingroup\let\do\@makeother\dospecials
  \catcode`\{=1 \catcode`\}=2 \doi@aux}
\providecommand{\doi@aux}[1]{\endgroup\texttt{#1}}
\makeatother
\providecommand*\mcitethebibliography{\thebibliography}
\csname @ifundefined\endcsname{endmcitethebibliography}
  {\let\endmcitethebibliography\endthebibliography}{}

\end{document}